# Imaging electric field dynamics with graphene optoelectronics

Jason Horng[‡,1,2], Halleh B. Balch[‡,1,2], Allister F. McGuire[3], Hsin-Zon Tsai[1], Patrick Forrester[1], Michael F. Crommie[1,2,4], Bianxiao Cui[3], & Feng Wang[1,2,4,*]

[1] Department of Physics, University of California Berkeley, Berkeley, California 94720, USA.
[2] Kavli Energy NanoSciences Institute at the University of California Berkeley and the Lawrence Berkeley National Laboratory, Berkeley, California 94720, USA.
[3] Department of Chemistry, Stanford University Stanford, California 94305, USA
[4] Materials Sciences Division, Lawrence Berkeley National Laboratory, Berkeley, California 94720, USA.
[‡] These authors contributed equally to this work.
[*] Correspondence should be addressed to F.W. (fengwang76@berkeley.edu)

**The use of electric fields for signaling and control in liquids is widespread, spanning bioelectric activity in cells[1,2] to electrical manipulation of microstructures in lab-on-a-chip devices[3–6]. However, an appropriate tool to resolve the spatio-temporal distribution of electric fields over a large dynamic range has yet to be developed. Here, we present a label-free method to image local electric fields in real time and under ambient conditions. Our technique combines the unique gate-variable optical transitions of graphene with a critically-coupled planar waveguide platform that enables highly-sensitive detection of local electric fields with a voltage sensitivity of a few microvolts, a spatial resolution of tens of micrometers and a frequency response over tens of kHz. Our new imaging platform enables parallel detection of electric fields over a large field of view and can be tailored to broad applications spanning lab-on-a-chip device engineering to analysis of bioelectric phenomena.**

Signaling and manipulation through the control of electric field distributions is ubiquitous to both biological and physical systems. For example, intercellular electrical activity is central to the signaling and computation of excitable cells such as cardiac and neuronal cells.[7] The voltages generated by bio-electric fields span three orders of magnitude and can fluctuate on the scale of microseconds to hours.[8,9] Likewise, the generation and manipulation of electric fields inside microscopic channels is the backbone of microfluidics and lab-on-a-chip diagnostics.[10,11] In the latter applications, electric field gradients are intentionally designed to create strong dielectrophoretic forces, which permit the trapping and control of individual chemical droplets or biological samples along programmable pathways covering hundreds of microns.[3,4,11]

Over the past decade, there has been a concerted effort to develop new techniques to observe and analyze the dynamic field fluctuations in liquids, such as the development of multitransistorarrays[12,13], voltage-sensitive dyes[14–18], and new computational paradigms to model[4,19,20] electric field behavior. However, it remains an outstanding challenge to achieve label-free, non-perturbative detection with high field sensitivity and high spatio-temporal resolution. For example, the study of network-scale biological activity requires the ability to non-perturbatively record single-cell signals over a large field of view and with sub-millisecond temporal sensitivity. The predominant label-free method of detecting local electric fields across cellular networks are multielectrode and multitransistor arrays.[19,21] These techniques require prefabricated electrode arrays that are difficult to scale up, are individually amplified, and do not provide the flexibility to measure electrical fields at arbitrary spatial positions. Compared to electrical techniques, optical imaging permits high throughput detection that is compatible with simultaneous complementary measurements. Despite these advantages, there is little research into label-free optical platforms that permit the spatio-temporal detection of electric field[22] distributions. Here, we address this subset of problems with a critically coupled waveguide-amplified graphene electric field (CAGE) imaging platform, which permits label-free imaging of the dynamics of electric fields in solutions under non-equilibrium conditions.

## Results

**Design the CAGE imaging platform**

Atomically thin materials like graphene offer exceptional opportunities for electric field sensing due to their unique physical properties and intrinsic sensitivity to their environment. Over the past decade, graphene's optoelectronic properties have been studied extensively.[23,24] In addition to its distinctive electronic transport properties, graphene couples strongly to light across the visible and the infrared frequency range. A pristine sheet of graphene has a constant absorption of $\pi\alpha$ ~2.3% at all frequencies.[25] In addition, this absorption can be modulated through electrostatic gating: a gate-induced shift of the Fermi energy that forbids specific optical transitions in graphene due to Pauli blocking (Figure 1a). We employ this field-dependent optical absorption to achieve highly sensitive and parallel optical detection of local electrical fields.

Realizing this goal requires new optical designs that optimize the detection sensitivity and parallel readout simultaneously. For example, simple transmission mode imaging of gated monolayer graphene yields a 1% change of transmitted light over a gate voltage change of 200mV around the Pauli blocking region. This results in a voltage sensitivity of only 2mV, accounting for a typical laser with noise levels of $10^{-4}$. However, physical and biological applications require orders of magnitude improvement in voltage sensitivity. To increase the light-matter interaction in graphene, researchers have explored different approaches such as coupling graphene to silicon waveguides[26–28], photonic crystal cavities[28–30], plasmon resonances[31–33], and metamaterials[33,34]. These existing methods can improve the voltage sensitivity of optical detection, but are incompatible with parallel detection and imaging.

Here, we demonstrate highly sensitive, high-speed optical imaging of local electric field dynamics in solutions using graphene and a critical-coupled planar waveguide. Using a custom numerical simulation, we designed the critically-coupled waveguide amplified platform to obtain the so-called critical coupling condition[35] where the effective absorption of monolayer graphene approaches 100%. Close to the critical-coupling point, the voltage sensitivity can be enhanced by orders of magnitude. In addition, the planar waveguide permits two-dimensional time-resolved imaging of the electric field distribution in the solution above graphene. This critically coupled waveguide-amplified graphene electric field (CAGE) imaging achieves a voltage sensitivity down to a few microvolts, a response speed of microseconds, spatial resolution of several microns, and highly parallel readout of the dynamics of electrical field distributions.

Figure 1b schematically illustrates the structure of the CAGE imaging platform. Large-area graphene is grown by chemical vapor deposition and transferred to a prism coated with 150nm of $Ta_2O_5$ ($n$ = 2.0856), forming a high refractive index waveguide. We use an equilateral SCHOTT SF-11 glass (SF-11) prism ($n$ = 1.743) with a 1000 nm top layer of $SiO_2$ ($n$ = 1.444) to evanescently couple a collimated 1.55um beam to and from the CAGE sensor. The reflected light is then collected by an infrared objective and imaged onto an InGaAs camera and photodiode.

The concept of waveguide critical coupling is illustrated in Figure 1c. A collimated s-polarized 1.55 μm incident beam is coupled into the transverse electric (TE) mode of the waveguide at an oblique angle so that condition for total internal reflection is satisfied at the SF-11/$SiO_2$ and $Ta_2O_5$/solution interfaces. Under this framework, we can describe the waveguide as a Fabry-Perot cavity with high reflection coefficients $|r_1|$ and $|r_2|$ at the two interfaces, where $|r_1|$ and $|r_2|$ deviate from unity due to the frustrated total internal reflection from the finite $SiO_2$ thickness at the first interface and the absorption of graphene at the second interface. At resonance coupling, the total reflection R from the Fabry-Perot cavity is described by $R = \frac{(|r_1|-|r_2|)^2}{(1-|r_1||r_2|)^2}$ (Supplementary Note I). The value of $|r_2|$ can be varied *in situ* by electrostatic gating of graphene. To attain the highest sensitivity to local electric fields, we gate the graphene to a specific Fermi energy that generates the largest relative change in optical reflection for a given local electric field. At the critical-coupling condition, $|r_1| = |r_2|$ and the total reflection, $R$, equals zero, at which point all of the light is absorbed by graphene. Consequently, close to the critical coupling condition, the light-graphene interaction is strongly enhanced and the background reflection is very weak contributing to the high voltage sensitivity of CAGE imaging.

**High sensitivity via critical coupling**

Figure 2a shows the gate-dependent reflectivity response of a typical CAGE detector in a saline solution (blue solid line). The incident 1.55 μm beam is collimated and TE-polarized. The resonance condition of the waveguide determines the angle of incident light inside the SF-11 coupling prism to be ~ 60 degrees from normal. With a 1000 nm $SiO_2$ layer, the frustrated total internal reflection $|r_1| = 98.2\%$. The reflection coefficient at the $Ta_2O_5$/solution $|r_2|$ is 97.5% at the charge neutral point ($V_g$=-0.14V) due to graphene absorption, which leads to a total reflection of $R$~ 1.5%. Graphene absorption can be set subsequently by electrostatic gating through the solution. The critical coupling condition $|r_1| = |r_2|$ is realized at $V_g$=+0.41V (electron doped) and $V_g$ = -0.69V (hole doped), resulting the lowest total reflectivity, R. (The residue value of R = 0.63% at critical coupling is due in part to a slight divergence in the incident beam and in part due to defects in the CVD-grown graphene and the waveguide thin film deposition.) Further increase of carrier doping leads to a decrease of graphene absorption corresponding to $|r_1| < |r_2|$ and one obtains an increase in total reflectivity. The grey dashed line in Figure 2a shows the simulated optical response of our device using graphene absorption determined experimentally. (See Supplementary Note 2 and Supplementary Figure 2 for graphene absorption and Supplementary Note 3 and Supplementary Figure 3 for simulation details)

The sensitivity of CAGE detection to dynamics of the local electric field is characterized by the relative reflectivity change *dR/R* caused by a change in voltage induced by the local electric field, *dV*. Figure 2b shows the calculated *(dR/R)/dV* as a function of the gate voltage from Figure 2a. CAGE detection is most sensitive close to the critical coupling condition, where the greatest optical modulation is achieved for a given change in local electric field. For example, a single millivolt of field-induced voltage produces a 1.2% change in reflection at $V_g$= 0.53 V. This sensitivity is over 200 times higher than that of the direct transmission configuration (~1% optical change per 200 mV, Supplementary Note 2).The noise of the commercially available 15 mW 1.55 μm butterfly diode lasers is around 0.01% RMS across a 10 Hz-10 kHz bandwidth. Consequently, the CAGE detection permits optical readout of electrical voltages smaller than 10 μV across a wide field of view.

**Noise floor and bandwidth**

CAGE optical detection of small electric field fluctuations is demonstrated in Figure 3a. The gate voltage at graphene was set at the highest sensitivity point (0.53 V). We applied periodic rectangular pulse sequences with peak voltages ranging from 500 μV to 100 μV and recorded the optical readout with a wide band-pass filter from 10 Hz to 10 kHz. The relative reflectivity change is 0.58%, 0.23%, and 0.11% for 500 μV, 200 μV, and 100 μV voltage pulses, respectively. The results are consistent with the *(dR/R)/dV* = 1.2% per mV reported in Figure 2b. Clear periodic modulation is observed at $V_{pp}$ = 100μV with a signal to noise ratio (SNR) of 6.5. This measurement reflects optical detection capabilities of ~15 μV or, equivalently, a change of carrier concentration of $2.3 \times 10^8$ electrons per cm$^2$ in graphene. The noise level at 0.017% in Fig. 3a is due to a combination of laser intensity fluctuations and vibrations of optical components. Much higher SNR and therefore higher voltage sensitivity would be possible by improving the optomechanical and laser source stabilities.

The temporal response of CAGE detection is determined by the RC constant of the system, where C is the capacitance of the graphene/electrolyte interface and *R* is governed by graphene conductance. For local electrical field fluctuations, the effective area and capacitance is small and the response speeds can be very high. Experimentally, we characterized the frequency response of the CAGE detector using a relatively large graphene area (200 μm by 400 μm), underscoring the applicability of our measurements to large fields of view. Figure 3b shows that the *dR/R* value remains largely constant up to 10 kHz and decreases at higher frequencies. These measurements yield a 3dB frequency of 13 kHz (Figure 3b dashed) and an RC constant of 20 μs. This 10 kHz fast response enables direct observation of electric field dynamics on timescales spanning action potentials[2,3] and electrophoretic manipulation[8,9].

**Electric field imaging**

Figure 4 demonstrates the capability of CAGE imaging to spatially resolve electric field dynamics. Figure 4a shows a schematic of our experimental setup. A platinum/iridium microelectrode is placed 5 μm above the device to create a spatially varying electric field distribution. The spatially resolved reflection from the graphene plane is projected to the image plane using a long working distance near-IR objective, which we image onto an InGaAs camera.

The spatio-temporal dynamics of local electric fields in solution are captured by the CAGE device at the critical coupling condition and imaged onto a 1D InGaAs camera array in Figures 4b and 4c. A 10 mV electrical pulse with a 200 ms duration is applied at the microelectrode (red waveform Figure 4b) generating a 1.8 mV local potential at the graphene/solution interface beneath the microelectrode tip (Supplementary Note 4).

The temporal response of the local electric field as a function of position are given as time traces in Figure 4b. Positions A1, A2, and A3, are increasingly distant from the field source. At position A1, immediately below the excitation electrode, we observe the perturbed reflection intensity due to a fast transient voltage peak that decays in ~25 ms to a stationary potential (solid yellow line). As one moves increasingly distant from the field source, the fast transient peak correspondingly reduces (green and blue solid lines). This behavior matches well with the results of our finite element simulation and may be qualitatively described by the equivalent circuit shown in Supplementary Note 4. This equivalent circuit consists of the solution resistance, impedance from the electrode/solution interface, and the impedance from the graphene/solution interface. The results of the calculation using the equivalent circuit are shown as dashed lines in Figure 4b. The voltage sensitivity is determined by the ~100 µV RMS noise present before the onset of the excitation; this sets the upper bound of voltage resolution in the 1D camera array.

The complete data set of the local field described above is shown in Figure 4c. We observe the field dynamics with 5 ms temporal resolution and 100 µV voltage sensitivity across a full 200 µm. The lower voltage sensitivity in the imaging mode compared with a single InGaAs photodiode is due to the slower speed and a limited dynamic range of our InGaAs array. In the future, the voltage sensitivity and temporal resolution could be extended by adopting a better array detector.

Frames from a CAGE video of local electric field dynamics recorded via an 80 Hz two-dimensional (2D) InGaAs camera is shown in Figure 5. The spatial resolution of our device is on the order of 10 µm (Supplementary Note 5 and Supplementary Figure 5). Each frame is normalized by an image taken in zero-field. Compared to the recordings in Figure 4b, we expect that time t = -10 ms to t = 40 ms captures the period in which a positive transient voltage emerges and subsequently dissipates away from the excitation microelectrode, and t = 190 ms to t = 240 ms captures the period in which a negative transient voltage recovers to zero. Indeed, the data show that a positive voltage emerges and then diffuses spatially in frames 1–4, and a negative voltage appears and recovers to the equilibrium state in frames 5–8 (Supplementary Note 4). These results demonstrate that dynamic spatial variations of local electric fields can be imaged in real time via the CAGE imaging platform.

**Discussion**

In summary, we present a method of imaging local electric field dynamics under ambient conditions with high voltage and spatio-temporal resolution through the critically-coupled waveguide amplified graphene electric field (CAGE) imaging platform. This label-free and highly parallel technique offers over 200-fold improvement over conventional graphene based optical sensing and resolves sub-15 µV fluctuations with a bandwidth of 10 kHz across a wide field of view. The CAGE imaging platform is capable of operating under a wide range of chemical and thermal conditions, may be used simultaneously with complementary measurements, and may be spectrally tailored to enable broad applications from improved engineering of lab-on-a-chip devices to sensing bioelectric phenomena across cellular networks.

## Methods
**Sample preparation**
The CAGE imaging structure consists of 1000 nm $SiO_2$ (coupling layer) and 150 nm $Ta_2O_5$ (waveguide layer) deposited on one face of a 1-cm equilateral SF-11 glass prism by ion-assisted deposition. The structure was designed using a custom Python simulation (Supplementary Note 3 and Supplementary Figure 3) and fabricated by Edmund Optics. A large area graphene film was grown on copper foil using chemical vapor deposition (CVD). A 1-$cm^2$ area graphene was transferred onto the waveguide surface by PMMA-supported transfer. We obtain high-quality large-area graphene with near-uniform optical absorption at the device/solution interface. The Pt (2 nm)/Au (60 nm) electrodes were deposited on the graphene to make electrical contacts. The metal electrodes were insulated with nitrocellulose lacquer to prevent water-Au chemical reactions during measurements. The device was mounted in a solution chamber printed from PR48 resin (Autodesk) which permits access to the device from both the top and bottom. All data were obtained in a saline solution (155 mM NaCl, 2.966 mM $Na_2HPO_4$, 1.0588 mM $KH_2PO_4$) except for imaging data (Figure 4) obtained in 1mg $L^{-1}$ NaCl in water to accommodate the camera's frame rate. An external gate voltage $V_g$(0.53V) was applied through a Ag/AgCl electrode in solution to set the Fermi energy of graphene and to test the optical response of CAGE detection under electrostatic gating. For the spatially resolved measurements, an external gate voltage $V_g$ (1.1V) was applied through the microelectrode, whose high impedance at the electrode/solution interface requires a larger applied voltage but yields the same voltage bias and critical coupling condition at the detector's graphene/solution interface. A 10 mV electrical pulse generates a 1.8 mV local field at the graphene/solution interface beneath the Pt/Ir microelectrode tip (World Precision Instruments, #PTM23B05KTH) (Supplementary Figure 4b). To this we applied a small modulation to the micro-positioned microelectrode insulated in parylene with only the final 2 μm exposed to the solution.

**Optical measurements**
Supplementary figure 1c shows, in detail, the optical set-up used in the study. A stable, 1.55 um, 15 mW laser beam is generated by a butterfly telecomm laser (Newport Model 708 8-Channel Butterfly) with a current and temperature controller (Newport Model 9016 Modular Controller). The polarization is tuned to the TE-direction by a half-wave plate and further cleaned by a calcite polarizer. In the imaging mode, the incident beam is collimated and coupled into the CAGE platform for optimal sensitivity. In the scanning detection mode, the incident beam is controlled with a 17.5 cm focusing lens to have a numerical aperture of 0.002 and selects an area at the graphene interface. The incident light couples into the waveguide from one side of the prism. The prism coated with the planar waveguide is placed on a XY-translational stage and a rotating stage which allow for fine-tuning of the sample position and incident angle. The reflected light is then collected by a 10X MPlan objective and sent into an InGaAs two-dimensional camera (Allied Vision Technologies Goldeye 008 SWIR), an InGaAs one-dimensional camera (Andor 1.7μm InGaAs DU490A) and into a low-noise InGaAs photodetector, respectively. A circular iris is used to select the probing area for photodiode measurements.

**Data availability**

The data that support the findings of this study are available from the corresponding author upon request.

**References:**
1. Plonsey, R. (Duke U. & Barr, R. C. *Bioelectricity A Quantitative Approach*. (Springer, 2002).
2. Cohen, A. E. & Venkatachalam, V. Bringing bioelectricity to light. *Annu. Rev. Biophys.* **43,** 211–32 (2014).
3. Gagnon, Z. R. Cellular dielectrophoresis: Applications to the characterization, manipulation, separation and patterning of cells. *Electrophoresis* **32,** 2466–2487 (2011).
4. Çetin, B. & Li, D. Dielectrophoresis in microfluidics technology. *Electrophoresis* **32,** 2410–2427 (2011).
5. Link, D. R. *et al.* Electric Control of Droplets in Microfluidic Devices. *Angew. Chemie Int. Ed.* **45,** 2556–2560 (2006).
6. Hunt, T. P., Issadore, D. & Westervelt, R. M. Integrated circuit/microfluidic chip to programmably trap and move cells and droplets with dielectrophoresis. *Lab Chip* **8,** 81–87 (2008).
7. Yuste, R. From the neuron doctrine to neural networks. *Nat. Rev. Neurosci.* **16,** 487–497 (2015).
8. Xie, C., Lin, Z., Hanson, L., Cui, Y. & Cui, B. Intracellular recording of action potentials by nanopillar electroporation. *Nat. Nanotechnol.* **7,** 185–90 (2012).
9. Buzsáki, G., Anastassiou, C. a & Koch, C. The origin of extracellular fields and currents--EEG, ECoG, LFP and spikes. *Nat. Rev. Neurosci.* **13,** 407–20 (2012).
10. Dorfman, K. D., King, S. B., Olson, D. W., Thomas, J. D. P. & Tree, D. R. Beyond Gel Electrophoresis: Microfluidic Separations, Fluorescence Burst Analysis, and DNA Stretching. *Chem. Rev.* **113,** 2584–2667 (2013).
11. Zhang, C., Khoshmanesh, K., Mitchell, A. & Kalantar-zadeh, K. Dielectrophoresis for manipulation of micro/nano particles in microfluidic systems. *Anal. Bioanal. Chem.* **396,** 401–420 (2010).
12. Lambacher, A. *et al.* Electrical imaging of neuronal activity by multi-transistor-array (MTA) recording at 7.8 μm resolution. *Appl. Phys. A* **79,** 1607–1611 (2004).
13. Hutzler, M. *et al.* High-Resolution Multitransistor Array Recording of Electrical Field Potentials in Cultured Brain Slices. *J. Neurophysiol.* **96,** 1638–1645 (2006).
14. Lavis, L. D. & Raines, R. T. Bright ideas for chemical biology. *ACS Chem. Biol.* **3,** 142–55 (2008).
15. Li, L. S. Fluorescence probes for membrane potentials based on mesoscopic electron transfer. *Nano Lett.* **7,** 2981–2986 (2007).
16. Park, J. *et al.* Screening Fluorescent Voltage Indicators with Spontaneously Spiking HEK Cells. *PLoS One* **8,** e85221 (2013).

17. Miller, E. W. *et al.* Optically monitoring voltage in neurons by photo-induced electron transfer through molecular wires. *Proc. Natl. Acad. Sci. U. S. A.* **109,** 2114–9 (2012).
18. KW Park, *et al.* Single Molecule Quantum-Confined Stark Effect Measurements of Semiconductor Nanoparticles at Room Temperature. *ACS Nano* **6(11)** 10013–10023 (2012).
19. Einevoll, G. T., Kayser, C., Logothetis, N. K. & Panzeri, S. Modelling and analysis of local field potentials for studying the function of cortical circuits. *Nat Rev Neurosci* **14,** 770–785 (2013).
20. Angle, M. R., Cui, B. & Melosh, N. A. Nanotechnology and neurophysiology. *Large-Scale Rec. Technol.* **32,** 132–140 (2015).
21. Spira, M. E. & Hai, A. Multi-electrode array technologies for neuroscience and cardiology. *Nat Nano* **8,** 83–94 (2013).
22. Wang, Y., Shan, X., Wang, S. & Tao, N. Imaging Local Electric Field Distribution by Plasmonic Impedance Microscopy. (2016).
23. Nair, R. R. *et al.* Fine Structure Constant Defines Visual Transparency of Graphene. *Science (80-. ).* **320,** 1308–1308 (2008).
24. Wang, F. *et al.* Gate-variable optical transitions in graphene. *Science* **320,** 206–209 (2008).
25. Mak, K. F., Ju, L., Wang, F. & Heinz, T. F. Optical spectroscopy of graphene: From the far infrared to the ultraviolet. *Solid State Commun.* **152,** 1341–1349 (2012).
26. Liu, M. *et al.* A graphene-based broadband optical modulator. *Nature* **474,** 64–67 (2011).
27. Koester, S. J. & Li, M. Waveguide-Coupled Graphene Optoelectronics. *Sel. Top. Quantum Electron. IEEE J.* **20,** 84–94 (2014).
28. Majumdar, A., Kim, J., Vuckovic, J. & Wang, F. Electrical Control of Silicon Photonic Crystal Cavity by Graphene. *Nano Lett.* **13,** 515–518 (2013).
29. Shiue, R.-J. *et al.* Enhanced photodetection in graphene-integrated photonic crystal cavity. *Appl. Phys. Lett.* **103,** (2013).
30. Gan, X. *et al.* Strong enhancement of light-matter interaction in graphene coupled to a photonic crystal nanocavity. *Nano Lett.* **12,** 5626 (2012).
31. Grigorenko, A. N., Polini, M. & Novoselov, K. S. Graphene plasmonics. *Nat Phot.* **6,** 749–758 (2012).
32. Chen, J. *et al.* Optical nano-imaging of gate-tunable graphene plasmons. *Nature* 3–7 (2012).
33. Fei, Z. *et al.* Gate-tuning of graphene plasmons revealed by infrared nano-imaging. *Nature* **487,** 82–85 (2012).
34. Lee, S. H., Choi, J., Kim, H.-D., Choi, H. & Min, B. Ultrafast refractive index control of a terahertz graphene metamaterial. *Sci. Rep.* **3,** 2135 (2013).
35. Liu, Y. *et al.* Approaching total absorption at near infrared in a large area monolayer graphene by critical coupling. *Appl. Phys. Lett.* **105,** (2014).

**Acknowledgements**

The authors thank Felix Alfonso, Hui-Ling Han, Sufei Shi and Zhiwen Shi for helpful discussions and thank Andreas Bastian and the Ember Team at Autodesk for help designing and printing the device mount and solution holder. This work was supported by the National Science Foundation grant DMR- 1344302 (optical measurements, simulations, device fabrication), and by the U.S. Department of Energy Office of Basic Energy Sciences contract no. DE-AC02-05CH11231 Nanomachine program (graphene fabrication). F.W. and B.C. acknowledge support from the David and Lucile Packard fellowship. H.B.B. acknowledges support from the NSF Graduate Research Fellowship (grant DGE 1106400). A.F.M. acknowledges support from the Stanford Bio-X Graduate Fellowship.

**Author Contributions**

F.W. and B.C. conceived of the experiment. J.H. and H.B.B. contributed equally to this work. J.H., H.B.B., and F.W. designed the experiment. H-Z.T., P.F. and M.C. grew high-quality CVD graphene. J.H. and H.B.B. developed the simulations, fabricated the devices, built the optical setup, performed the optical measurements, and wrote the manuscript. J.H., H.B.B., and A.F.M. analyzed the data. All authors contributed to the discussion of the results and to the manuscript.

**How to cite this article:** Horng, J. and Balch, HB. *et al.* Imaging electric field dynamics with graphene optoelectronics. *Nat Commun.*

**Competing Financial Interests**

The authors declare no competing financial interests.

# Figure 1

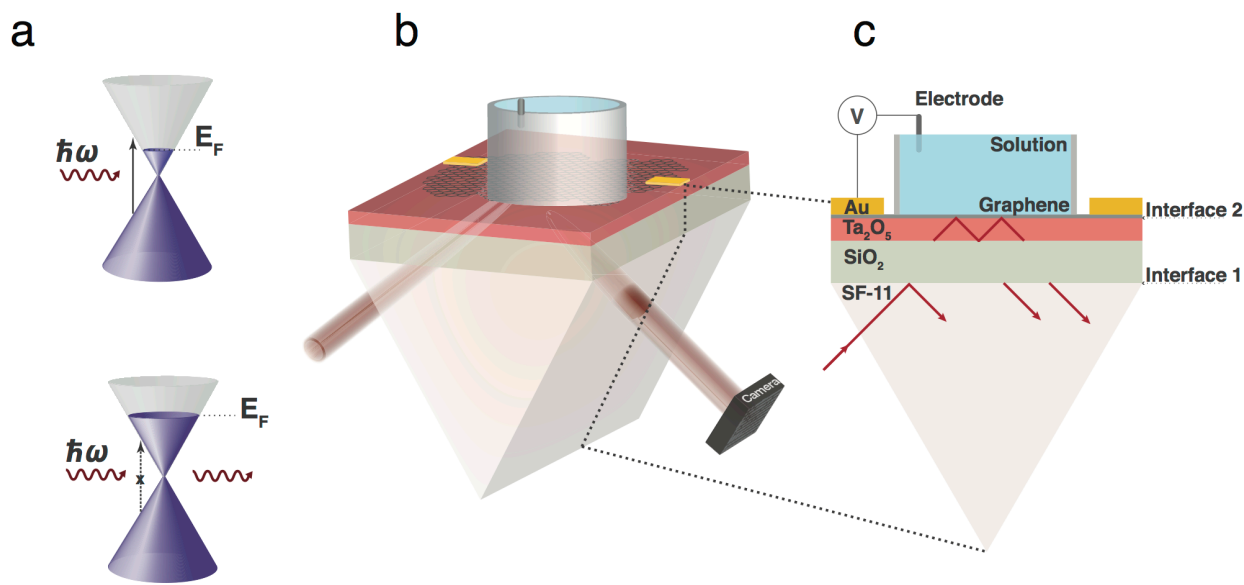

**Figure 1 | Graphene optoelectronics and CAGE imaging device. a,** Graphene interband transitions wherein $E_F$ shifts upon gating. The optical modulation is strongest for electronic transitions near the Fermi surface, $2|E_F| = \hbar\omega$, wherein the presence of an external field can shift the Fermi energy and prohibit optical absorption due to Pauli blocking (right). **b,** Critically coupled waveguide-amplified graphene electric field (CAGE) imaging device in which a transverse-electric (TE) polarized collimated incident beam at 1.55 $\mu$m is coupled through the prism coupler (prism and green layer) into the waveguide (red). Through the waveguide, the beam probes the graphene/solution interface (grey). The critical coupling condition is achieved by varying the incident light angle and by electrostatic gating of graphene through the saline solution (blue). The out-coupled signal is detected by an InGaAs photodiode and/or camera. **c,** Cross section of the CAGE imaging platform. Interface 1 and Interface 2 form a Fabry-Perot cavity in which we obtain the critical coupling condition. The waveguide-amplified critical coupling condition sets the ratio of light coupling into the waveguide and light absorption at the graphene interface to unity. The optical contrast of local field fluctuations is maximized close to the critical coupling condition, which permits localized and sensitive electric field detection.

# Figure 2

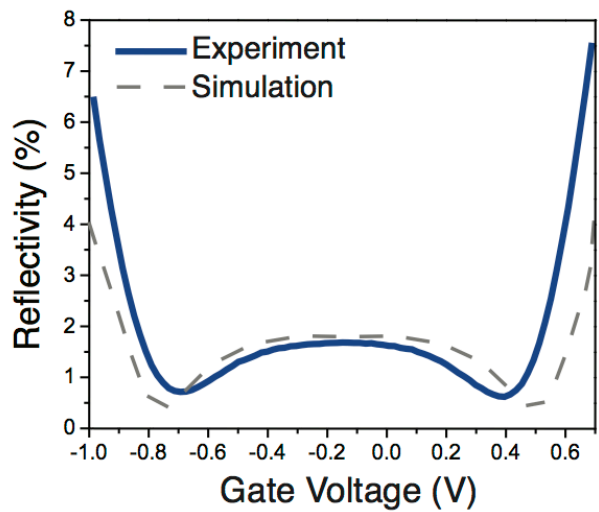 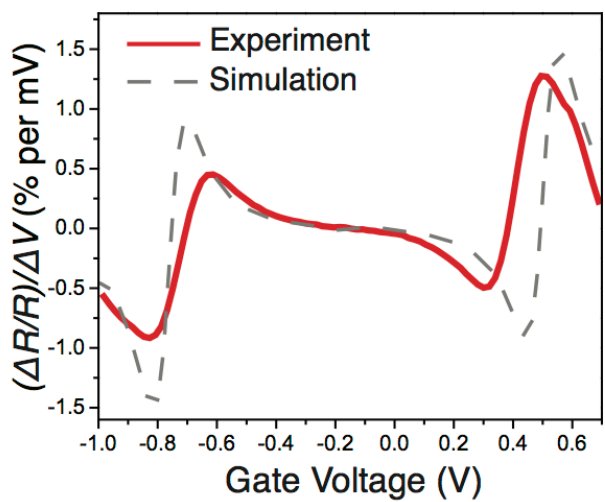

**Figure 2 | Optical response characterization of the CAGE sensor. a,** Gate-dependent optical reflectivity of the TE-polarized collimated 1.55μm beam incident at the waveguide resonance angle. The dip in the optical signal corresponds to the critical coupling condition. The grey dashed line shows the optical response expected from simulation (see text and Supplementary Note 2 for details). **b,** CAGE sensor voltage sensitivity, *(dR/R)/dV*, is derived from (a) for both experiment (red) and simulation (grey dashed line). We observe a maximum voltage sensitivity of 1.2% optical change per mV at $V_g$=+0.53V.

# Figure 3

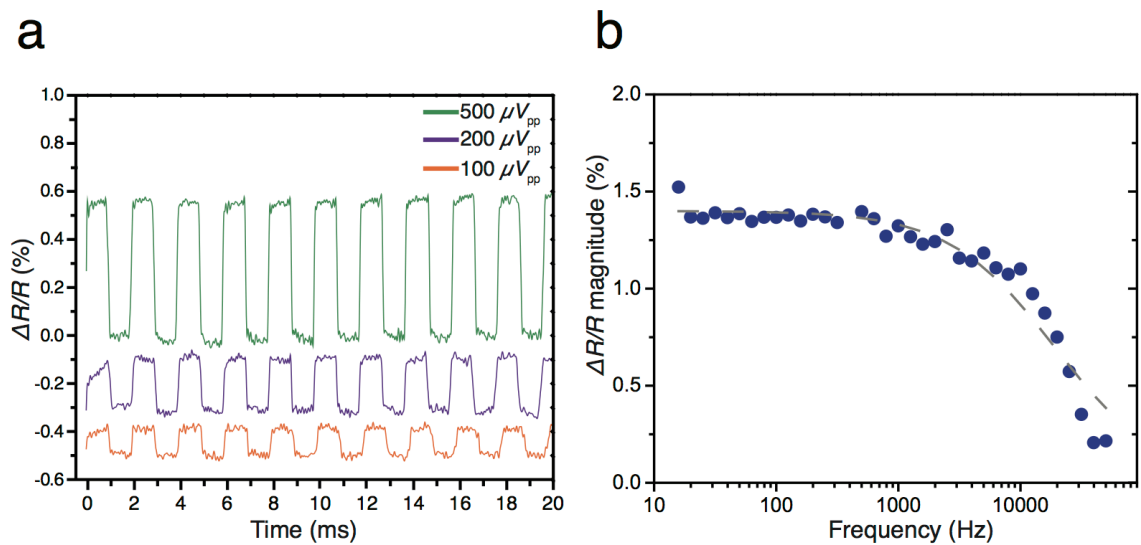

**Figure 3 | Voltage sensitivity and temporal bandwidth. a,** Optical detection of an applied modulating voltage. A periodic rectangular waveform is applied with peak-to-peak voltages of 500 $\mu V_{pp}$ (green), 200 $\mu V_{pp}$ (purple), and 100 $\mu V_{pp}$ (orange) with a 10 Hz-10 kHz bandpass filter. The optical response from the 100μV applied modulation demonstrates a SNR of 6.5 corresponding to an experimental detection limit of 15 μV. **b,** Frequency dependence of the optical signal demonstrating sensitivity to high-speed fluctuations up to 10 kHz. A 1 $mV_{pp}$ sinusoidal waveform with frequencies spanning 20 Hz to 50 kHz is applied. Shown here for a device with large-area (80,000 μm$^2$) graphene. The frequency bandwidth will increase inversely with graphene area as a consequence of graphene's high conductivity. The results provided by the equivalent circuit, accounting for the double layer capacitance and graphene resistance, is plotted as the grey dashed line.

# Figure 4

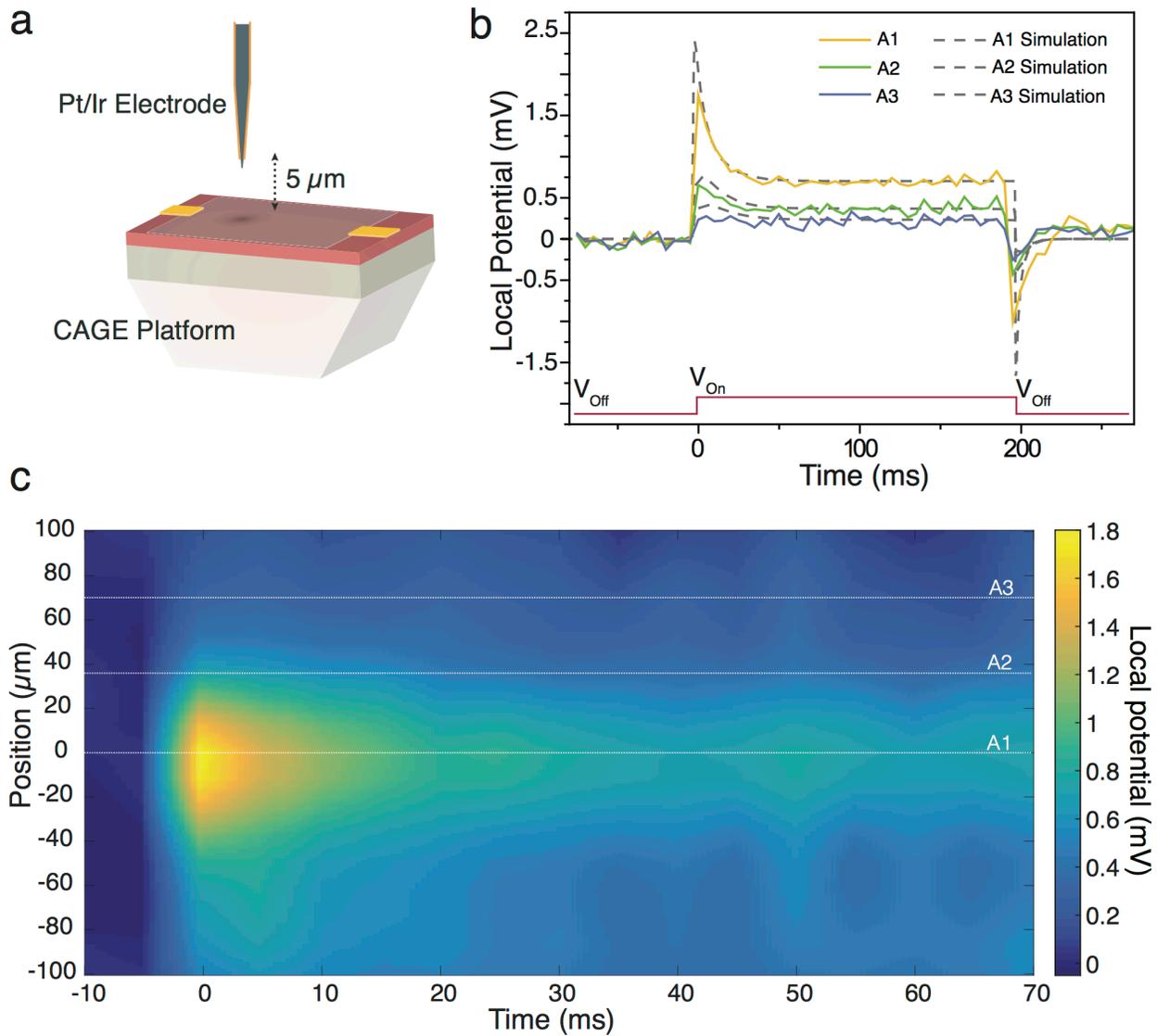

**Figure 4| Detecting local electric field fluctuations with spatial and temporal resolution. a,** Geometry of the experimental setup in which a waveform is applied to a 2 μm platinum/iridium microelectrode placed in solution 5 μm above the graphene surface of the detector. Applying a waveform to the microelectrode localizes the electric field and permits observation of the local electric field modulation in space and time. **b,** Temporal dynamics of the experimental (solid) and simulated (dashed) optical CAGE detection of the local electric field at different distances from the local potential source. The local field is generated by a 10 mV 200 ms pulse (red) applied to the microelectrode. The spatial location of A1 (yellow), A2 (green), and A3 (blue) is articulated by white dashed lines in 4c. **c,** CAGE image with spatio-temporal resolution of the first 70 ms of local electric field dynamics described in (b) projected onto a 1D 193 Hz InGaAs camera. The spatially resolved recording obtains ~100 μV sensitivity with 5ms temporal resolution.

**Figure 5**

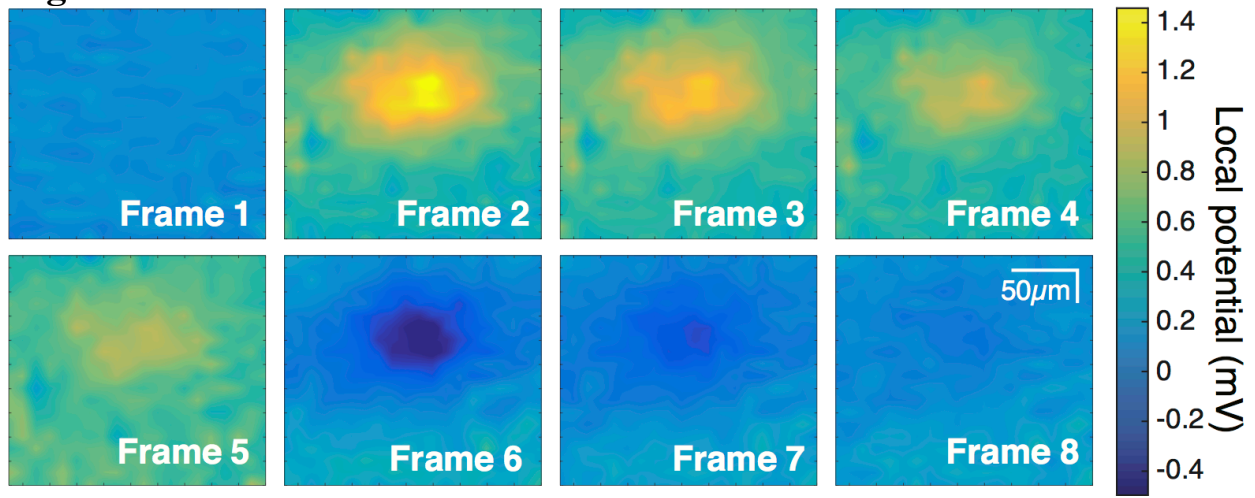

Still images from the single-shot recording in response to the same pulse in Figure 4 projected onto a 2D 80 Hz InGaAs camera. Frames 1-4 capture the first 50 ms of the field and its spatial diffusion throughout the solution while frames 5-8 begin at t =190 ms and capture the completion of the pulse and its recovery to equilibrium. Frames are separated by 12.5 ms. The reflection intensity in the stills is normalized to that without stimulation.